# Qunits: queried units for database search

Arnab Nandi
Computer Science, EECS
University of Michigan, Ann Arbor
arnab@umich.edu

H.V. Jagadish
Computer Science, EECS
University of Michigan, Ann Arbor
jag@umich.edu

## ABSTRACT

Keyword search against structured databases has become a popular topic of investigation, since many users find structured queries too hard to express, and enjoy the freedom of a "Google-like" query box into which search terms can be entered. Attempts to address this problem face a fundamental dilemma. Database querying is based on the logic of predicate evaluation, with a precisely defined answer set for a given query. On the other hand, in an information retrieval approach, ranked query results have long been accepted as far superior to results based on boolean query evaluation. As a consequence, when keyword queries are attempted against databases, relatively ad-hoc ranking mechanisms are invented (if ranking is used at all), and there is little leverage from the large body of IR literature regarding how to rank query results.

Our proposal is to create a clear separation between ranking and database querying. This divides the problem into two parts, and allows us to address these separately. The first task is to represent the database, conceptually, as a collection of independent "queried units", or *qunits*, each of which represents the desired result for some query against the database. The second task is to evaluate keyword queries against a collection of qunits, which can be treated as independent documents for query purposes, thereby permitting the use of standard IR techniques. We provide insights that encourage the use of this query paradigm, and discuss preliminary investigations into the efficacy of a qunits-based framework based on a prototype implementation.

## 1. INTRODUCTION

Keyword search in databases has received great attention in the recent past, to accommodate queries from users who do not know the structure of the database or are not able to write a formal structured query. Projects such as BANKS [3], Discover [13], ObjectRank [2] and XRank [10]

Supported in part by NSF grants IIS-0438909 and IIS-0741620, and by a Yahoo! Research Fellowship.

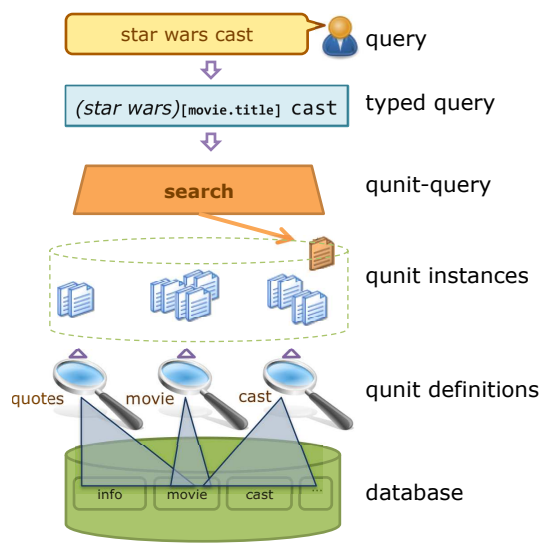

**Figure 1: The Qunits Search Paradigm : The database is translated into a collection of independent *qunits*, which can be treated as documents for standard IR-like document retrieval.**

have all focused on providing better algorithms for ranking elements in a database given a keyword query. These algorithms use various data models for the purposes of scoring results. For example, BANKS considers candidate results that form a minimal spanning tree of the keywords. ObjectRank, on the other hand, combines tuple-level PageRank from a pre-computed data graph with keyword matching. A common theme across such efforts is the great attention paid to different ranking models for search.

This paradigm is similar to one adopted by the information retrieval community who strive to improve document search. For them, the task is to create an algorithm to find the best *document* given a keyword query. Various approaches, such as TF/IDF and PageRank, have emerged as popular approaches to solving the search problem, where the user is presented with a list of *snippets* from top-ranking documents.

However, unlike document collections, large schemas do not have the luxury of a clear definition of *what* a document is, or what a search result comprises. Even if one assumes that the ranking of the results is correct, it remains unclear what information should be presented to the user, i.e. what snippets from a database search result. Records in a schema-rich database have many possible sub-records, parent records, and joinable records, each of which is a candidate for inclusion in the result, but only a few of which are potentially useful to the average user.

For example, consider the keyword query *george clooney movies*. When issued on a standard document search engine, the challenge is straightforward: pick the best document corresponding to the query and return it. However, for a database, the same task becomes nontrivial. In the case of a relational database, should we return just singular tuples that contains all the words from the query? Do we perform some joins to denormalize the result before presenting it to the user? If yes, how many tables do we include, and how much information do we present from them? How do we determine what the important attributes are? The title of a movie isn't unique due to remakes and sequels, and hence it will not be a key. How do we determine it is more useful to the user than the primary key, *movie_id*? Clearly, there are no obvious answers to these questions.

In other words, a database search system requires not just a mechanism to find matches and rank them, but also a systematic method to demarcate the extent of the results. It may appear that simply including the complete spanning tree of joins used to perform the match (in the case of relational systems, such as BANKS), or including the complete sub-tree rooted at the least common ancestor of matching nodes (in the case of XML systems, such as XSearch), will suffice. However these simple techniques are insufficient, often including both too much unwanted information and too little desired information.

One cause of too much information is the chaining of joins through unimportant references, which happens frequently in normalized databases. One reason for too little information, particularly in relational systems, is the prevalence of `id` attributes due to normalization. If this were the only issue, it could be addressed by performing a value join every time an internal "id" element is encountered in the result. However, the use of simple rules such as this also appears insufficient.

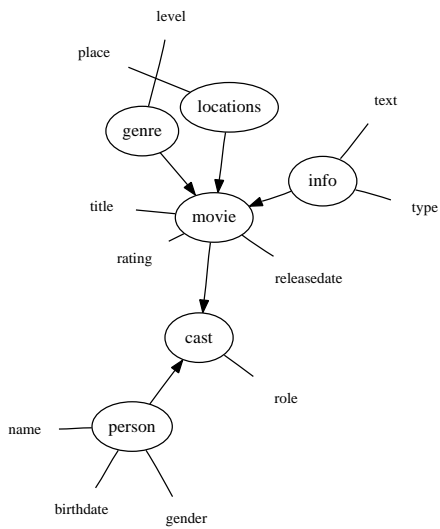

**Figure 2: A simplified example schema**

Consider, as shown in Fig. 2, part of the example schema for the Internet Movie Database. The schema comprises primarily of two entities, *person* and *movie*. Additional tables, such as *cast*, establish relationships, while tables such as *genre* are used to normalize common strings. Given a keyword query such as *george clooney movies*, a natural solution is to perform a join of ``name = george clooney`` from the **person** table, with the **movies** table via **cast**. The **movies** table is normalized and contains `id` pointers to the **locations**, **genre**, and **info** table. There is nothing in terms of database structure to distinguish between these three references through `id`. Yet it is probably the case that resolving the id references to the shooting locations and genre is desired, but not to include the lengthy plot outline (located in the **info** table) in the search result.

In addition to the challenge of demarcating an actual result, the problem of identifying which result to return for a query still remains. Based on prior work in the area of search query log analysis [19, 29], and analyses described in Sections 5.1 and 5.2, we assert that:

- A majority of users' queries are underspecified

- The mapping between true information need and the actual query terms is frequently not one-to-one

- The challenge of mapping a query to a specific fragment of the database is nontrivial

In this paper, we introduce the notion of a "qunit", representing a quantified unit of information in response to a user's query. The search problem then becomes one of choosing the most appropriate qunit(s) to return, in ranked order – a problem that is much cleaner to address than the traditional approach of first finding a match in the database and then determining how much to return. The fundamentals of this concept are described in Sec. 2. Sec. 3 describes search inside a qunits based system, while Sec. 4 discusses methods presents different techniques for identifying qunits in a database.

We study the nature of search behavior, and then experimentally evaluate the quality of search results from our qunits-based approaches in Sec. 5. We close with a review of related work in Sec. 6 and some concluding remarks in Sec. 7.

## 2. QUNIT AS A CONCEPTUAL UNIT

We now define the notions of a *qunit* and *qunit utility*:

**Qunit:** A *qunit* is the basic, independent semantic unit of information in a database.

Every user has a "mental map" of how a database is organized. Qunits are an embodiment of this perceived organization. The organization may not be the most efficient representation, in terms of storage, or other computational considerations. For example, while the user may like to think of the IMDb as a collection of actor profiles and movie listings, the underlying relational database would store normalized "fact" tables and "entity" tables, for superior performance. In other words, a qunit can be considered as a view on a database, which we call the *base expression*, with an additional *conversion expression* that determines the presentation of the qunit.

Qunits do not need to be constrained by the underlying data representation model. This allows qunits to be much richer than views. For example, consider the IMDb database. We would like to create a qunit definition corresponding to the information need "cast". A cast is defined as the people associated with a movie, and rather than have the name of the movie repeated with each tuple, we may prefer to have a nested presentation with the movie title on top and one tuple for each cast member. The base data in IMDb is relational, and against its schema, we would write the *base expression* in SQL with the *conversion expression* in XSL-like markup as follows:

```
SELECT * FROM person, cast, movie
WHERE cast.movie_id = movie.id AND
      cast.person_id = person.id AND
      movie.title = "$x"
RETURN
   <cast movie="$x">
   <foreach:tuple>
       <person>$person.name</person>
   </foreach:tuple>
   </cast>
```

The combination of these two expressions forms our *qunit definition*. On applying this definition to a database, we derive qunit *instances*, one per movie.

**Qunit Utility:** The *utility* of a qunit is the importance of a qunit to a user query, in the context of the overall intuitive organization of the database. This applies to both qunit definitions and qunit instances.

The number of possible views in a database is very large. Hence, the total number of *candidate* qunit definitions is an inordinately large number. The utility score of a qunit is required for selecting a top-ranking set of *useful* qunits from the large pool of candidate qunits. From a user's perspective, this process captures the most salient and relevant concepts that should be returned for a particular query.

It should be noted that qunit utility is a subjective metric. It is dependent on the intent and nature of the user, and is not necessarily uniform for all users. For our purpose, we approximate this metric with clearly defined, objective surrogates. This is similar in spirit to measuring *document relevance* in information retrieval, where TF/IDF is a well-defined and widely accepted objective metric used to approximate document relevance.

Once qunits have been defined, we will model the database as a flat collection of independent qunits. Qunits may overlap, and in fact one qunit may even completely include another. However, these overlaps will not be given any special attention. Similarly, references (links, joins, etc) are expected to have been resolved to the extent desired at the time that qunits were defined. Thus in our movie database, a qunit such as `movie` can list its actors, but there is no explicit link between a `movie` qunit and an `actor` qunit. In subsequent querying, each qunit is treated as an independent entity.

Note, however, that context information, not part of the qunit presented to the user, may often be useful for purposes of search and ranking. For example, link structures may be used for network analysis. Our model explicitly allows for this.

## 3. QUNIT-BASED SEARCH

Consider the user query, *star wars cast*, as shown in Fig.1. Queries are first processed to identify entities using standard query segmentation techniques[30].

In our case one high-ranking segmentation is "[movie.name] [cast]" and this has a very high overlap with the qunit definition that involves a join between "movie.name" and "cast". Now, standard IR techniques can be used to evaluate this query against qunit instances of the identified type; each considered independently even if they contain elements in common. The qunit instance describing the cast of the movie *Star Wars* is chosen as the appropriate result.

As we can see, the qunits based approach is a far cleaner approach to model database search. In current models of keyword search in databases, several heuristics are applied to leverage the database structure to construct a result on the fly. These heuristics are often based on the assumption that the structure within the database reflects the semantics assumed by the user (though data / link cardinality is not necessarily an evidence of importance), and that all structure is actually relevant towards ranking (though internal *id* fields are never really meant for search).

The benefit of maintaining a clear separation between ranking and database content is that structured information can be considered as one source of information amongst many others. This makes our system easier to extend and enhance with additional IR methods for ranking, such as relevance feedback. Additionally, it allows us to concentrate on making the database more efficient using indices and query optimization, without having to worry about extraneous issues such as search and ranking.

It is important to note that this conceptual demarcation of rankings and results does not imply materialization of all qunits. In fact, there is no requirement that qunits be materialized, and we expect that most qunits will not be materialized in most implementations.

## 4. QUNIT DERIVATION

In the previous section, we discussed how to execute a query in a database that already had qunits identified. One possibility is for the database creator to identify qunits manually at the time of database creation. Since the subject matter expert is likely to have the best knowledge of the data in the database, such expert human qunit identification is likely to be superior to anything that automated techniques can provide. Note that identifying qunits merely involves writing a set of view definitions for commonly expected query result types and the manual effort involved is likely to be only a small part of the total cost of database design. If a forms-based database interface has been designed, the set of possible returned results constitute a good human-specified set of qunits.

Even though manual expert identification of qunits is the best, it may not always be feasible. Certainly, legacy systems have already been created without qunits being created. As such, automated techniques for finding qunits in a database are important. In this section, we discuss the automated derivation of qunits from a database.

Given a database, there are several possible sources of information that can be used to infer qunits. Of course, there is the database schema. Since identifying qunits requires us to write base expressions defining them, and writing these expressions requires schema knowledge, the use of schema knowledge is a starting point. In addition, there are three independent possible sources of information worth considering. The first and most obvious is the data contained in the database itself. A second source of information is the history of keyword queries posed to the system from previous instances of search on the database, also known as a query log. The final source of information about the database is the source of evidence outside the database, such as published results and reports that could be based on information from the database in question, or a similar data source. We consider each of these three sources in turn.

### 4.1 Using Schema and Data

Inspired by previous efforts[2, 18] that model the database as a graph, we utilize the concept of *queriability* of a schema described in [15] to infer the important schema entities and attributes in a database. Queriability is defined as the likelihood of a schema element to be used in a query, and is computed using the cardinality of the data that the schema represents. Qunit base expressions are generated by looking at the top-$k$ schema entities based on descending queriability score. Each of the top-$k_1$ schema entities is then expanded to include the top-$k_2$ neighboring entities as a join, where $k_1$ and $k_2$ are tunable parameters. While this method is intuitive and self-contained within the database, there are many instances where using just the data as an arbiter for qunit derivation is suboptimal. Specifically, in the case of underspecified queries such as "george clooney", with the example schema in Fig. 2, creating a qunit for "person" would result in the inclusion of important movie "genre" and the unimportant movie "location" tables, since every movie has a genre and location. However, while there are many users who are interested in "george clooney" as an actor in romantic comedies, there is very little interest in the locations he has been filmed at.

### 4.2 Using Query Logs

We use a *query rollup* strategy for query logs, inspired by the observation that keyword queries are inherently underspecified, and hence the qunit definition for an under-specified query is an aggregation of the qunit definitions of its specializations. For example, if the expected qunit for the query "george clooney" is a personality profile about the actor George Clooney, it can be constructed by considering the popular specialized variations of this query, such as "george clooney actor", "george clooney movies", and so on.

We begin by sampling the database for entities, and look them up in the search query log. The found query log entries are then collected, along with the number of times they occur in the query log (query frequency). For each query, we then map each recognized entity on to the schema, constructing simple join plans. We then consider the popular plan fragments for the qunit definitions.

As an example, consider the schema element "person.name". Instances of this element are "george clooney" and "tom hanks", which are looked up in the query log, where we find search queries such as "george clooney actor", "george clooney batman" and "tom hanks castaway". Additionally, searches for "movie.name" instances, "batman" and "castaway" and "cast.role" instance "actor" would also lead to the same 3 queries. Given these three queries, we have hence built an annotated set of schema links, where "person.name" links to "cast.role" once, and to "movie.name" twice. This suggests that the rollup of the qunit representing "person.name" should contain "movie.name" and "cast.role", in that order.

### 4.3 Using External Evidence

There is a wealth of useful information that exists in the form of *external evidence*. We now propose a third method that uses external evidence to create qunits for the database. External evidence can be in the form of existing "reports" – published results of queries to the database, or relevant web pages that present parts of the data. Such evidence is common in an enterprise setting where such reports and web pages may be published and exchanged but the queries to the data are not published. By considering each piece of evidence as a qunit instance, the goal is to learn qunit definitions.

In our running example, our movie search engine is aware of the large amount of relevant organized information on the Internet. Movie information from sources such as the Wikipedia [1] are well organized and popular. Since the actual information in Wikipedia will have a great overlap with that from IMDb, our goal is thus to learn the organization of this overlapped data from Wikipedia.

By using methods similar to query rollup, we use records in the database to identify entities in documents. We then compute "signatures" for each web page, utilizing the DOM tree and frequency of each occurrence. An example of a type signature for the cast page for a movie on Wikipedia would be ((person.name:1) (movie.name:40)), which would suggest using person.name as a label field, followed by a "foreach" consisting of movie.name, based on the relative cardinality in the signature and the number of the tuples generated in our qunit base expression. By aggregating the type signatures over a collection of pages, we can infer the appropriate qunit definition.

## 5. EXPERIMENTS AND EVALUATION

In this section, we begin by first discussing a brief pre-experiment to explore the nature of keyword searches that real users posed against a structured database. The following subsections describe the use of a real-world querylog to evaluate the efficacy of qunit based methods.

### 5.1 Understanding Search

The Internet Movie Database or IMDb [2] is a well-known repository of movie-related information on the Internet. To gain a deeper understanding of search behavior, we performed a user study with five users, all familiar with IMDb and all with a moderate interest in movies. The subjects had a large variance in knowledge about databases. Two were graduate students specializing in databases, while the other three were non-computer science majors. The subjects were asked to consider a hypothetical movie database that could answer all movie-related queries. Given this, the users were asked to come up with five "information needs", and the corresponding keyword queries that they would use to query the database.

---

[1] http://wikipedia.org
[2] http://imdb.com

As expected, nearly all our subjects look for the summary page of a movie. But this information need is expressed in five different ways by the users. Similarly, the cast of a movie and finding connections between two actors are also common interests. Once again these are expressed in many different ways. Conversely, a query that only specifies the title of the movie may be specified on account of four different information needs (first column of the table). Similarly, users may specify an actor's name when they mean to look for either of two different pieces of information – the actor's filmography, or information about co-actors. Clearly, there exists a many-to-many relationship between information need and queries.

| info. need | keyword query | [title] | [title] box office | [actor] | [award] [year] | [actor] [actor] | [genre] | [title] ost | [title] cast | [title] [freetext] | movie [freetext] | [title] year | [title] posters | [title] plot | don't know |
|---|---|---|---|---|---|---|---|---|---|---|---|---|---|---|---|
| movie summary | a,c | | | | | | | | | a | d | b | | d | |
| cast | | e | | | | | | | c,d | | | | | | |
| filmography | | | | e,a | | | | | | | | | | | |
| coactorship | | | | | | a, c | | b | | | | | | | |
| posters | | | | | | | | | | | | | b | | |
| related movies | e | | | | | | | | | | | | | | |
| awards | | | | | a | | | | | | | | | | |
| movies of period | | | | | | | | | | | | | | | b |
| charts / lists | | | | | | | | e | | | | d | | | |
| recommendations | | | | | | | | | | | | b | | | e |
| soundtracks | | | | | | | | c | | | | | | | |
| trivia | c | | | | | | | | | | | | | | |
| box office | | | d | | | | | | | | | | | | |

**Table 1: Information Needs vs Keyword Queries. Five users (a, b, c, d, e) were each asked for their movie-related information needs, and what queries they would use to search for them.**

Another key observation is that 10 of the 25 queries here are single entity queries, 8 of which are *underspecified* – the query could be written better by adding additional predicates. This happened even though we reassured users that we *could* answer all movie-related queries.

The results of our interviews are displayed in Table 1. Each row in this table is an information need suggested by one or more users. Each column is the query structure the user thought to use to obtain an answer for this information need. The users themselves stated specific examples; what we are presenting is the conceptual abstraction. For example if the user said they would query for *"star wars cast"*, we abstract it to query type `[title] cast`. Notice that the unmatched portion of the query(*cast*) is still relevant to the schema structure and is hence considered an attribute. Conversely, users often issue queries with words that are non-structural details about the result, such as "movie space transponders". We consider these words free-form text in our query analysis.

Entries in this matrix identify data points from individual subjects in our experiment. Note that some users came up with multiple queries to satisfy the same information need, and hence are entered more than once in the corresponding rows.

## 5.2 Movie Querylog Benchmark

To construct a typical workload, we use a real world dataset of web search engine query logs [26] spanning 650K users and 20M queries. All query strings are first aggregated to combine all identities into a single anonymous crowd, and only queries that resulted in a navigation to the `www.imdb.com` domain are considered, resulting in 98,549 queries, or 46,901 unique queries. We consider this to be our *base query* log for the IMDb dataset. Additionally, we were able to identify movie-related terms in about 93% of the unique queries (calculated by sampling).

We then construct a benchmark query log by first classifying the base query log into various types. Tokens in the query log are first replaced with schema "types" by looking for the largest possible string overlaps with entities in the database. This leaves us with typed templates, such as "[name] movies" for "george clooney movies". We then randomly pick two queries that match each of the top (by frequency) 14 templates, giving us 28 queries that we use as a workload for qualitative assessment.

We observed that our dataset reflects properties consistent with previous reports on query logs. At least 36% of the distinct queries to the search engine were "single entity queries" that were just the name of an actor, or the title of a movie, while 20% were "entity attribute queries", such as "terminator cast". Approximately 2% of the queries contained more than one entity such as "angelina jolie tombraider", while less than 2% of the queries contained a complex query structure involving aggregate functions such as "highest box office revenue".

## 5.3 Evaluating Result Quality

The result quality of a search system is measured by its ability to satisfy a user's information need. This metric is subjective due to diversity of user intent and cannot be evaluated against a single hand-crafted gold standard. We conducted a result relevance study using a real-world search query log as described in the following subsection, against the Internet Movie Database. We asked 20 users to compare the results returned by each search algorithm, for 25 different search queries, rating each result between 1 (result is correct and relevant) and 0 (result is wrong or irrelevant).

For our experiments, we created a survey using 25 of the 28 queries from the movie querylog benchmark. The workload generated using the query log is first issued on each of the competing algorithms and their results are collected. For our algorithms mentioned in Sec. 4, we implement a prototype in Java, using the imdb.com database (converted using IMDbPy(`http://imdbpy.sf.net`) to 15 tables, 34M tuples) and the base query log as our derivation data. To avoid the influence of a presentation format on our results, all information was converted by hand into a paragraph in a simplified natural English language with short phrases. To remove bias, the phrases were collated from two independent sources. 20 users were then sourced using the Amazon Mechanical Turk (`http://mturk.com`) service, all being moderate to advanced users of search engines, with moderate to high interest in movies. Users were then primed with a sample "information need" and "query" combination : *need to find out more about julio iglesias* being the need, and *"julio iglesias"* being the search query term. Users were then presented with a set of possible answers from a search engine, and were asked to rate the answers presented with one of the options listed in Table 2.

Users were then asked to repeat this task for the 25 search queries mentioned above. The table also shows the score we internally assigned for each option. If the answer is incorrect or uninformative it obviously should be scored 0. If it is the correct answer, it obviously should be scored 1. Where an answer is partially correct(incomplete or excessive), we should give it a score between 0 and 1 depending on how correct it is. An average value for this is 0.5.

| score | rating |
|---|---|
| 0 | provides incorrect information |
| 0 | provides no information above the query |
| .5 | provides correct, but incomplete information |
| .5 | provides correct, but excessive information |
| 1.0 | provides correct information |

**Table 2: Survey Options**

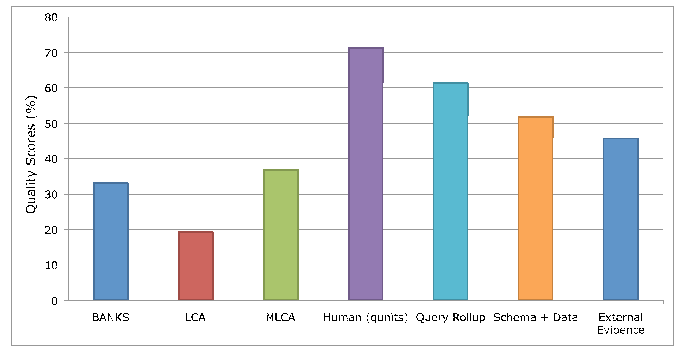

**Figure 3: Comparing Result Quality against Traditional Methods : "Human" indicates hand-generated qunits.**

To provide an objective example of a qunit-based system, we utilize the structure and layout of the imdb.com website as an expert-determined qunit set. Each page on the website is considered a unique qunit instance, identified by a unique URL format. A list of the qunits is generated by performing a breadth-first crawl starting at the homepage, of 100,000 pages of the website and clustering the different types of URLs. Qunit definitions were then created by hand based on each type of URL, and queried against the test workload. Users were observed to be in agreement with each other, with a third of the questions having an 80% or higher of majority for the winning answer.

We now compare the performance of currently available approaches against the qunits described in the derivation section, using a prototype based on ideas from Sec. 3. To do this, we first ran all the queries on the BANKS[3] online demonstration. A crawl of the imdb.com website was converted to XML to retrieve the LCA(Lowest Common Ancestor) and MLCA [20] (Meaningful Lowest Common Ancestor). The MLCA operator is unique in that it ensures that the LCA derived is unique to the combination of queried nodes that connect to it, improving result relevance. In addition to these algorithms, we also include a data point for the theoretical maximum performance in keyword search, where the user rates every search result from that algorithm as a perfect match.

Results are presented in Fig. 3 by considering the average relevance score for each algorithm across the query workload. As we can see, we are still quite far away from reaching the theoretical maximum for result quality. Yet, qunit-based querying clearly outperforms existing methods.

## 6. RELATED WORK

We acknowledge various bodies of relevant prior work in this section.

Previous efforts towards solving keyword search for databases have concentrated predominantly on ranking. Initial work such as BANKS [3, 18] and DBXplorer [1] use inverted indices on databases to return tuple joins in the form of spanning trees of matched query words and discuss efficient methods of deriving such trees. In the case of XML, one strategy is to identify the smallest element that contains some or all keywords [10], or to identify the smallest element that is *meaningful* [20]. Later work [2] in this area concentrates further on the ranking aspects of the problem, using a metric of authority transfer on a data graph to improve result quality. While these methods provide solutions towards the improvement of ranked results, the composition of the results themselves is not explored.

Work by Koutrika, Simitsis and Ionnadis on *Précis*[28]

receives special mention in this context. Précis responses to keyword queries are natural language compositions of "information directly related to the query selections" along with "information implicitly related to them in various ways". Results are not intended to be complete, and instead concentrate on being "comprehensive", akin to the idea of providing users with most probable next browse-points in a browsable database. Their contributions include a rich data model to express facts for the sake of representation, and semi-automated methods to infer these structures by way of annotating a schema graph with weights, providing detailed analysis on execution time and satisfaction of the user. We embrace and extend some of the methods involved, as described in Sec. 3.

A major challenge for Précis, and other keyword query systems for databases, is that new database-specific ranking algorithms have to be invented. These are often very complex, and yet tend not to have the quality of ranking algorithms in IR, where they have been studied much more intensively and for much longer. With the qunits proposal, our primary goal is to satisfy the users immediate information need. We address this problem by separating the problem of qunit identification from the problem of query evaluation and ranking. Our aim is to answer the user's query intent with an atomic unit of information, and not to guide the user to the relevant sections in the database. Unlike Precis results, qunits results are independent of data models. The qunits paradigm first partitions the database into actual pieces of information called qunits, making it amenable to standard IR techniques.

Our solution to this problem can be considered a method to organize information such that it is more amenable to a user's information needs. Work has been done in organizing information along individual perspectives using faceted search. Each "facet" is a distinct attribute that users can use to search and browse records. [6, 12] talk about design considerations for building faceted search systems. [31] approaches the faceted search problem by picking hierarchies based on user search patterns. [27] surveys many techniques of dynamic taxonomies and combines them. However, it is clear that faceted search is ideal for homogeneous data with multiple attribute constraints, and not a database with many different kinds of information in it.

Work has been done in building search mechanisms aware of the application logic layer above the database. [5] utilize the architecture patterns in application logic to better index dynamic web applications. [9] approaches this problem by exploiting exposed web service interfaces to provide a unified concept-search interface.

Applications of database-backed query answering are slowly appearing in mainstream search engines. These services have been sucessful at implementing template-based search, where a keyword query is identified and rerouted as a stuctured query to a database, answering questions such as "what is the population of china". With qunits, we propose to take this idea front and center, and make this the central paradigm for all search.

Managing complex databases have also been studied in the context of schema summarization [33], using characteristics of the schema. The schema clusters can then be queried upon [34] with low overhead. However, summarizing a schema does not take into account the nature of the query workload or the multiple ways to organize data. A related body of research in view selection [11, 24] attempts to select a set of views in the database that minimizes query execution time of a workload given resource costs, but does not consider the *utility* of the information provided by each view.

The broader concept of making databases more accessible using a presentation data model has been discussed by Jagadish et al. [14]. Work on exposing the database query interface using automatically generated forms has been discussed in [16, 17], while [25] uses autocompletion to provide the user with database-specific information. However, both methods do not consider the option of using clues from outside the database to improve the query process.

## 7. CONCLUSION

Keyword queries against structured data is a recognized hard problem. Many researchers in both the database and information retrieval communities have attacked this problem, and made considerable progress. Nonetheless, the incompatibility of two paradigms has remained: IR techniques are not designed to deal with structured inter-linked data, and database techniques are not designed to produce results for queries that are under-specified and vague. This paper has elegantly bridged this gap through the concept of a qunit.

Additionally, this paradigm allows both techiniques to co-exist under the same ranking model. Search engines have started to provide interesting ways [8, 32] to simultaneously search over a heterogenous collection of documents and structured data, with interesting implications for search relevance and ranking. Qunits provide a clean approach to solving this problem.

For the keyword query front end, the structured database is nothing more than a collection of independent qunits; so standard information retrieval techniques can be applied to choose the appropriate qunits, rank them, and return them to the user. For the structured database backend, each qunit is nothing more than a view definition, with specific instance tuples in the view being computed on demand; there is no issue of underspecified or vague queries.

In this paper, we presented multiple techniques to derive the qunits for a structured database and determine their utility for various query types; we presented an algorithm to evaluate keyword queries against such a database of qunits, based on typifying the query; we experimentally evaluated these algorithms, and showed that users find qunit-based query results substantially superior to results from the best keyword-based database query systems available today.

We believe that the concept of a qunit is powerful, and that this is merely a first paper introducing this idea. In future work, we expect to be able to substantially improve upon the qunit finding and utility assignment algorithms presented above; we expect to deal with qunit evolution over time as user interests mutate during the life of a database system; we expect to extend qunit notions to databases with substantial mixed text content and to use IR techniques to query the text content in conjunction with the database structure.